\documentclass[twocolumn,aps,showpacs]{revtex4}

\newtheorem{theorem}{Theorem}
\newtheorem{prop}{Proposition}

\newcommand{\mo}{\mu_0}
\newcommand{\mot}{\frac{\mo}{3}}
\newcommand{\be}{\begin{equation}}
\newcommand{\ee}{\end{equation}}
\usepackage{pstricks}
\usepackage{pst-coil}
\usepackage{amssymb}
\usepackage{graphics,epsf}
\begin{document}
\title[Singularities in spherical dust collapse]{Geometry and topology of singularities in
spherical dust collapse}

\author{Brien C Nolan (1)\footnote{e-mail: brien.nolan@dcu.ie} \& Filipe C Mena
(2,3)\footnote{e-mail: fmena@math.uminho.pt}}

\affiliation{(1)\ School of Mathematical Sciences, Dublin City
University, Glasnevin, Dublin 9, Ireland.}

\affiliation{(2)\ Departamento de Matem\'{a}tica, Universidade do
Minho, Gualtar, 4710 Braga, Portugal} \affiliation{(3)\ School of
Mathematical Sciences, Queen Mary, University of London, London E1
4NS, UK.}

\begin{abstract}
We derive some more results on the nature of the singularities
arising in the collapse of inhomogeneous dust spheres. (i) It is
shown that there are future-pointing radial and non-radial
time-like geodesics emerging from the singularity if and only if
there are future-pointing radial null geodesics emerging from the
singularity. (ii) Limits of various space-time invariants and
other useful quantities (relating to Thorne's
point-cigar-barrel-pancake classification and to isotropy/entropy
measures) are studied in the approach to the singularity. (iii)
The topology of the singularity is studied from the point of view
of ideal boundary structure. In each case, the different nature of
the visible and censored region of the singularity is emphasized.

\end{abstract}



\maketitle

\section{Introduction and summary}
The role played by regular initial data in determining the final
state of marginally bound
spherically symmetric collapse of inhomogeneous dust is
well understood. A comprehensive
description of this role was first given in \cite{sj} where explicit
necessary and sufficient conditions on the initial data were derived
for the existence of a naked singularity. A naked singularity here
means a singular region from which there emanates a future-pointing
(f.p.) causal geodesic. The result referred to dealt with the
existence of f.p.\ radial null geodesics.
Motivated by a desire to
(i) study the stability of this result with respect to the
introduction of angular momentum and (ii) generalise, in the
relevant subsets of the initial data space,
to all f.p.\ causal geodesics, we have studied non-radial null
geodesics in spherical dust collapse \cite{mn1}. We showed that
there exist f.p.\ non-radial null geodesics emanating from the
singularity if and only if there exist f.p.\ radial null geodesics
emanating from the singularity. It would be useful to know whether
this result can be extended to timelike geodesics in order to give
a complete account of all possible causal geodesics. Interesting
results in this direction were recently given in \cite{desh}.

The existence of non-radial geodesics outgoing from the singularity
might suggest that, topologically, the singularity is not a point.
It is therefore interesting to study in more detail the structure of the
singularity and in particular to investigate its topological properties.
This is a difficult problem since, firstly, the singularity
is not part of the space-time and there is no unique definition for
the space-time boundary (see \cite{seno} for a review). Secondly,
given the complexity of the geodesic equations, it is difficult
to study the topology of a given singular boundary.

In order to have a first intuition about the
singularity `shape', we can start by studying
the behaviour of the expansion of dust world lines
in the neighbourhood of the singularity. In this way,
we can classify the singularities
as points, cigars, barrels or pancakes \cite{dsincosmo}.

However, this classification says little about topological issues.
For example, in a dust filled Friedmann-Lemaitre-Robertson-Walker
universe the big bang singularity is point-like in this
classification, but is foliated by 2-spheres according to a scheme
used below. Furthermore, there are simple examples where this
classification scheme does not give any results such as for the
singularity in Schwarzschild space-time with negative mass. We
recall that in the context of the self-similar collapse of a
scalar field, Christodoulou \cite{christo-scalar} has found an
example a singular boundary (see below) which is foliated by
points in some sectors and by spheres in other sectors. So, a
space-time singularity might not have a distinct unique topology
and can possess complicated geometrical features (see also e.g.\
the Curzon singularity \cite{scott}). It is therefore of
importance to investigate some of these non-trivial issues in the
context of spherical dust collapse.

This paper aims to study carefully some aspects of the singularity
and its topological properties in Lemaitre-Tolman (LT)
space-times. In section II we briefly describe the marginally
bound LT model and recall previous results of \cite{sj} and
\cite{mn1}. In section III, we consider the problem of extending
the result of \cite{mn1} to all f.p.\ causal geodesics. In section
IV, we investigate the causal structure of the singularity by
studying the behaviour of ingoing radial null geodesics. In
section V, we shall use Thorne's classification scheme, which
turns out to be of interest in this case beyond merely classifying
the singularity. We shall also study the behaviour of a number of
space-time invariants in order to investigate the structure of the
singularity. These invariants involve the shear and the space-time
curvature and have been used e.g.\ in studies of the Weyl
curvature hypothesis \cite{penrose} as well as in the context of
isotropic singularities \cite{gw}. In section VI, we address
topological issues by considering non-radial null geodesics
emerging from the singularity for self-similar solutions. We end
with some conclusions and comments. All our considerations are
restricted to the local visibility of the singularity. Another
important point to keep in mind is that when dealing with radial
geodesics, we will for the most part treat them as curves in the
Lorentzian 2-space $\theta=$constant, $\phi=$constant. So a
`unique' radial geodesic as referred to in e.g.\ Proposition 4 is
a unique curve in the $r-t$ plane, but is a 2-parameter family of
curves in the space-time. The exception is Section VI, where the
different angular parameters of radial curves play an essential
role.  A bullet $\bullet$ indicates the end of a proof. We use
$8\pi G=c=1$.
\section{Field equations and singularities}
We study spherical inhomogeneous dust collapse for the marginally
bound case. The line element is \cite{Lemaitre,Tolman,Bondi}
\begin{equation}
ds^2 = -dt^2 +(R_{,r}dr)^2 +R^2(r,t)d\Omega^2,\label{eq1}
\end{equation}
where $d\Omega^2$ is the line element for the unit 2-sphere.
We use subscripts to denote partial derivatives. The
Einstein field equations yield
\begin{eqnarray}
R_{,t}&=&-\sqrt{\frac{m(r)}{R}},\label{feq1}\\
\rho &=& \frac{m_{,r}}{R^2R_{,r}},\label{feq2}
\end{eqnarray}
the latter equation defining the density $\rho(r,t)$ of the dust;
$m(r)$ is arbitrary (but subject to certain conditions
introduced below) and will be referred to as the mass, although
in fact $m$ equals twice the Misner-Sharp mass. In particular,
this locates the apparent horizon at $R=m$. Solving (\ref{feq1})
gives
\begin{equation}
R^3(r,t)=\frac{9}{4}m(t_c(r)-t)^2,\label{req}
\end{equation}
where $t_c(r)$ is another arbitrary function which, in the process of collapse,
corresponds to the time of arrival of each shell $r$ to the singularity.

We assume that the collapse proceeds from a regular initial state;
i.e.\ at time $t=0$ (i) there are no singularities (all curvature
invariants are finite) and (ii) there are no trapped surfaces
i.e.\ $m(r)<R(r,0)$. We have the freedom of a coordinate rescaling
in (\ref{eq1}) which we can use to set $R(r,0)=r$. This gives
\begin{equation} t_c(r) = \frac{2}{3}\sqrt{\frac{r^3}{m}}.\label{feq3}
\end{equation}
Notice that this leaves just one arbitrary function, $m(r)$. There
is a curvature singularity called the shell-focussing singularity
along $t=t_c(r)$, so the ranges of the coordinates $r,t$ are
$0\leq r<\infty$ and $0\leq t<t_c(r)$. The possibility that the
shell-focussing singularity may be naked was first noted in
\cite{eardley-smarr}, and the first mathematical results on the
role of initial data were given in \cite{christo-dust}. We note
that only the subset $\{(r,t)=(0,t_0)\}$ (where $t_0:=t_c(0)$) of
this singularity may be visible \cite{christo-dust}. Assuming a
dust sphere of finite radius, we can restrict the range of $r$ to
$[0,b]$ for some $b>0$, and match it to a Schwarzschild exterior.
During the collapse of the dust sphere there can be also a
curvature singularity given by $R_{,r}=0$ along $t=t_{sc}(r)$,
where
\begin{equation}
t_{sc}(r) =2\frac{\sqrt{rm}}{m_{,r}}.
\end{equation}
This so-called shell-crossing singularity is gravitationally weak
\cite{newman,nolan99}, though what this means in terms of
continuation of the geometry is not yet known. Since we are
primarily interested in the shell-focussing singularity we impose
the condition that along each world-line $r=constant$, the
shell-crossing singularity does not precede the shell-focussing
singularity. That is, $t_c(r)<t_{sc}(r)$ for all $r>0$. This is
equivalent to taking $R_{,r}>0$ for all $r>0$, and yields
$rm_{,r}<3m, r>0$. We note that $t_{sc}(0)=t_0$.

The initial data consists of just one function $\mu(r):=\rho(r,0)$
defined on an interval $[0,b]$ where $b$ is the initial radius of
the collapsing dust sphere. The result of \cite{sj} confirmed in
\cite{mn1} is as follows:
\begin{theorem}
Let $b>0$ and $\mu\in C^3[0,b]$.  Let
\[ m(r)=\int_0^r x^2\mu(x)dx\]
defined on $[0,b]$ satisfy the no-shell crossing condition
$rm_{,r}<3m$ on $(0,b]$. Then the marginally bound collapse of the
dust sphere with initial radius $b$ and initial density profile
$\mu(r)$ results in a singularity from which there emanates a
f.p.\ radial null geodesic if and only if one of the following
conditions is satisfied.
\begin{enumerate}
\item $\mu^\prime(0)<0.$
\item $\mu^\prime(0)= 0$ and $\mu^{\prime\prime}(0)<0$.
\item $\mu^\prime(0)=\mu^{\prime\prime}(0)=0$ and
\[
\frac{\mu^{\prime\prime\prime}(0)}{(\mu_0)^{5/2}}\leq-\frac23(26\sqrt{3}+45)\simeq-60.0222,\]
where $\mu_0:=\mu(0)$.
\end{enumerate}
\end{theorem}
In \cite{mn1}, we found it convenient to make the following
definitions which will be used below.
\begin{eqnarray}
m(r)&=:&r^3(\mot+m_1),\qquad m_1(0)=0;\label{m1}\\
m_{,r}(r)&=:&r^2(\mo+m_2),\label{m2}\\
m_{,rr}(r)&=:&r(2\mo+m_3),\qquad m_3(0)=0.\label{m3}
\end{eqnarray}
Notice that $m_2(r)=\mu(r)-\mo$, and so $m_2\in C^3[0,b]$ and
$m_2(0)=0$. These terms are related as follows:
\begin{prop}
\label{hello} $m_1,m_3 \in C^3[0,b]$ and
\begin{eqnarray}
m_1^\prime(0)&=&\frac14 m_2^\prime(0),\qquad m_3^\prime(0)=3m_2^\prime(0);\label{p1eq1}\\
m_1^{\prime\prime}(0)&=&\frac15 m_2^{\prime\prime}(0),\qquad m_3^{\prime\prime}(0)=4m_2^{\prime\prime}(0);\label{p1eq2}\\
m_1^{\prime\prime\prime}(0)&=&\frac16
m_2^{\prime\prime\prime}(0),\qquad
m_3^{\prime\prime\prime}(0)=5m_2^{\prime\prime\prime}(0).\label{p1eq3}
\end{eqnarray}
\end{prop}

As we shall see below, the proofs of the results in \cite{mn1} sometimes require a
different approach in each of the following cases:
\begin{enumerate}
\renewcommand\labelenumi{\theenumi}
\renewcommand{\theenumi}{(\roman{enumi})}
\item $m_1^\prime(0)\neq 0.$
\item $m_1^\prime(0)=0, m_1^{\prime\prime}(0)\neq 0.$
\item $m_1^\prime(0)=0, m_1^{\prime\prime}(0)= 0, m_1^{\prime\prime\prime}(0)\neq 0.$
\end{enumerate}
\section{Time-like geodesics}
In this section, we determine the circumstances under which radial and
non-radial time-like geodesics may emanate from the singularity.

The governing equations for the  geodesics with angular momentum $L$ obtained
from the Euler-Lagrange equations are
\begin{eqnarray}
{\ddot t}+R_{,r}R_{,rt}\dot{r}^2+R_{,t}\frac{L^2}{R^3}&=&0,
\label{NRTLG1}\\
{R_{,r}}{\ddot r}+R_{,rr}{\dot r}^2+2R_{,rt}{\dot r}{\dot
t}-\frac{L^2}{R^3}&=&0,\label{NRTLG2}\\
-{\dot t}^2+(R_{,r}{\dot r})^2+\frac{L^2}{R^2}&=&-\epsilon, \label{NRTLG3}
\end{eqnarray}
where the over-dot represents differentiation with respect to an affine parameter $s$ along the
geodesics. We have $\epsilon=+1$ for time-like geodesics and $\epsilon=0$
for null geodesics. We are looking for
the existence of a solution of
(\ref{NRTLG1})-(\ref{NRTLG3}), which satisfies the following condition:
\\\\
{\bf Existence condition:}
\newline
There exists $\delta>0$ such that ${\dot t}$ is a non-negative,
integrable function of the affine parameter $s$ and ${\dot r}$ is
an integrable function of $s$ for $s\in[0,\delta)$ and such that
\[ \lim_{s\to 0^+} r(s) =0,\qquad \lim_{s\to 0^+} t(s) = t_0.\]

Non-negativity of ${\dot t}$ implies that the geodesic is
future-pointing. We shall now demonstrate the following result:
\begin{theorem}
Given regular initial data for marginally bound spherical dust
collapse subject to a condition which rules out shell-crossing
singularities, there are radial and non-radial timelike geodesics
which emanate
from the ensuing singularity if and only if there are radial null
geodesics which emanate from the singularity.
\end{theorem}
Proving one half of Theorem 2 is quite straightforward.
Considering the f.p.\ causal geodesics through a point $p$ of this
spherically symmetric space-time as curves in the $r-t$ plane, we
can compare slopes at $p$ using (\ref{NRTLG3}) to show that a
f.p.\ causal geodesic $\gamma$ through $p$, which is not the
unique outgoing radial null geodesic $\gamma_*$ through $p$,
locally precedes $\gamma_*$. That is if $\gamma,\gamma_*$ are
given by $t=t_\gamma(r)$, $t=t_{\gamma_*}(r)$ respectively with
$t_\gamma(r_0)=t_{\gamma_*}(r_0)$ where $r_0$ corresponds to $p$,
then $t_{\gamma_*}(r)>t_\gamma(r)$ for $r<r_0$. Hence if
$\gamma_*$ avoids the singularity i.e.\ reaches $r=0$ at some time
$t_*<t_c(0)$, then so too must $\gamma$. A general result of this
nature is given in \cite{nmg}. So we have
\begin{prop}
If there are no f.p.\ radial null geodesics emanating from the
singularity, then there are no f.p.\ causal geodesics emanating
from the singularity.
\end{prop}

Proving the converse result is more difficult. However it turns
out that the proof of Proposition 9 in \cite{mn1} relating to non-radial null
geodesics requires only minor modifications in order to be applied
in the present case. The idea of the proof is to identify a region
$\Omega$ bounded by a line $r=r_0>0$ and by curves $t=t_1(r),
t=t_2(r)$ which intersect at the singularity and out of which
time-like geodesics moving into the past may not pass. This
confinement is proven by carefully controlling the derivative of
$x/y$, where $x={\dot t}$ and $y=R_{,r}{\dot r}$, and the region
may be constructed only if there are radial null geodesics
emanating from the singularity. Using the geodesic equations
(\ref{NRTLG1})-(\ref{NRTLG3}), we
obtain
\begin{equation}
\frac{d}{ds}\left(\frac{x}{y}\right)
=\frac{L^2}{R^3y}\left[G\left(1+\epsilon\frac{R^2}{L^2}\right)+\sqrt{\frac{m}{R}}-\frac{x}{y}\right],
\label{xydot} \end{equation} where $G(r,t)=RR_{,rt}/R_{,r}$. The
definition of $\Omega$ guarantees that $G$ is positive therein.
The key part of the proof for null geodesics involves showing that
$d(x/y)/ds>0$ in $\Omega$. But positivity in the null case
($\epsilon=0$) clearly implies positivity in the time-like case
($\epsilon=+1$), and so the result of Proposition 9 of \cite{mn1}
carries through. Thus we can state:

\begin{prop}
Let $\mu(r)$ be such that there exist f.p.\ radial null geodesics
emanating from the singularity. Then there exist f.p.\ radial and
non-radial time-like geodesics emanating from the singularity.
\end{prop}
This completes the proof of Theorem 2 and, together with the results of
\cite{mn1}, gives a complete account
of the causal geodesics emanating from the shell-focussing singularity
in marginally bound spherical dust collapse.

In the next sections, we consider some
implications of the existence of this array of geodesics emerging from
the singularity.
\section{Structure of the singularity I: Causal structure}
One more item is required to determine the causal structure of
the singularity: the behaviour  of ingoing radial null geodesics (IRNG's).
This behaviour is the same for every case which gives rise to a
naked singularity and is summarised as follows:
\begin{prop}
Let $m$ be such that the initial data give rise to a naked
singularity. Then there exists a unique ingoing radial null
geodesic which terminates at $t=t_0, r=0$.
\end{prop}
\noindent{\bf Proof:} To see that such geodesics must exist,
consider an outgoing radial null geodesic $\gamma$ which emerges
from $r=0$ at some time $t_1<t_0$. Then the radial ingoing null
geodesics originating on $\gamma$ reach $r=0$ either before, at or
after $t=t_0$. Let $S_1$ be the set of values of $r$ on $\gamma$
for which the first outcome holds, and let $S_2$ be the set for
which the last outcome holds. Then any IRNG through points of
$\gamma$ for which $r$ satisfies $\sup\{S_1\}\leq r\leq
\inf\{S_2\}$ must terminate at $r=0, t=t_0$.

In order to prove uniqueness, let $\epsilon>0$ and consider two IRNG's
$t_1(r),t_2(r)$ terminating at $r=0,t=t_0$ and satisfying
$t_1(r)<t_2(r)$ on $(0,\epsilon)$. Note that the geodesics cannot intersect
in $t<t_c(r)$. The equation governing IRNG's is \begin{equation}
\frac{dt}{dr} = -R_{,r} =
-12^{-1/3}\frac{m_{,r}}{m^{2/3}}\frac{t_{sc}-t}{(t_c-t)^{1/3}},
\label{rreq}\end{equation}
and using $t_1(r)<t_2(r)<t_0$, we obtain
\[ \frac{d}{dr}(t_2-t_1)<
12^{-1/3}\frac{m_{,r}}{m^{2/3}}(t_c-t_0)^{-1/3}(t_2-t_1).\] If
$\mu^\prime(0)$ and $\mu^{\prime\prime}(0)$ are not both zero,
then the coefficient of $t_2-t_1$ on the right hand side is
integrable on $[0,\epsilon)$ and so $t_2(0)-t_1(0)=0$ gives
$t_2-t_1\equiv 0$. In the case where
$\mu^\prime(0)=\mu^{\prime\prime}(0)=0$,
$\mu^{\prime\prime\prime}(0)\neq0$ the transformation $u=r^3$,
$y=-\frac{72}{\mu^{\prime\prime\prime}(0)}\left(\frac{\mu_0}{3}\right)^{3/2}(t_0-t)$
casts the equation in a form such that exactly the same argument
may be used. $\bullet$

The fact that the visible portion of the singularity may be
identified with a point $(0,t_0)$ on the boundary of a local
coordinate chart means that this portion of the singularity must
be null. The fact that there is a unique ingoing radial null
geodesic terminating at this portion of the singularity indicates
that it has the form of an ingoing rather than an outgoing null
hypersurface. Thus the local causal structure of the singularity
is as indicated in Figure 1. The global structure of the
space-time depends on the initial radius of the dust sphere, which
will determine whether the singularity is locally or globally
naked. That is, if the initial radius of the dust
sphere is sufficiently small, then future pointing radial null
geodesics which emerge from the singularity will enter the vacuum
region prior to the formation of the apparent horizon. Since the
apparent horizon meets the event horizon at the boundary of the
star, such geodesics must therefore lie in the portion of
the Schwarzschild space-time outside the event horizon, and extend to
${\cal{J}}^+$, giving a globally naked singularity. There are also
cases where all f.p.\ causal geodesics emerging from the
singularity cross the apparent horizon and terminate at the future
singularity. In those cases the singularity is locally naked.

\begin{figure}[!htb]
\centerline{\def\epsfsize#1#2{1 #1}\epsffile{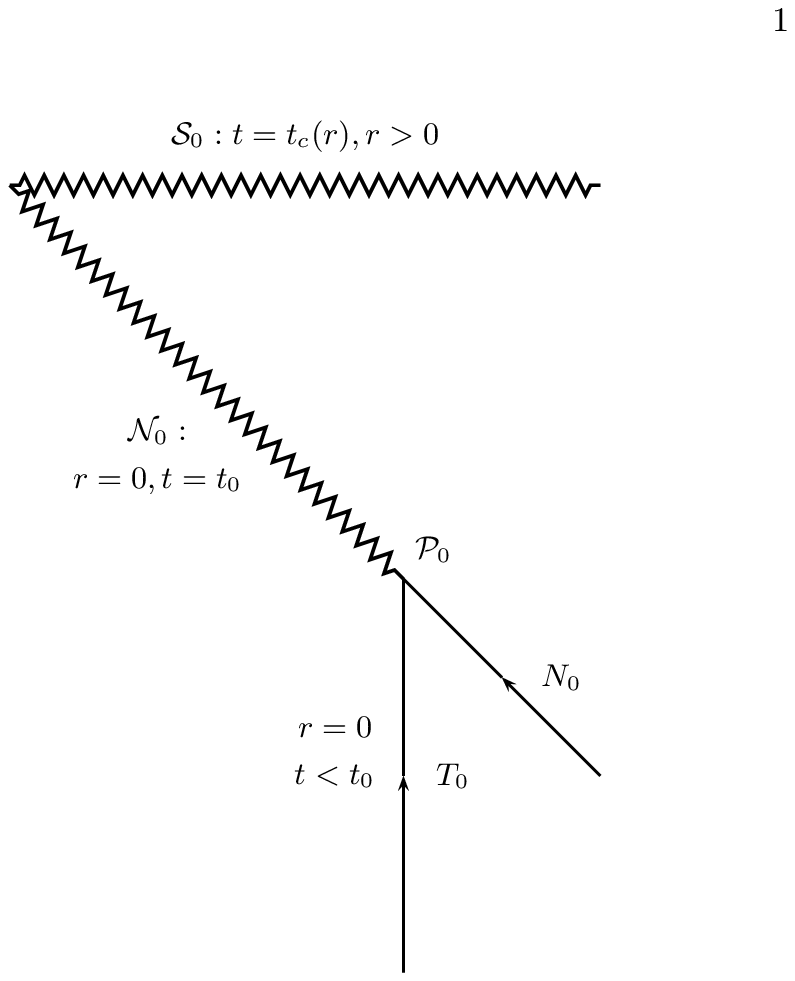}}
\caption{\label{figure1} Local causal structure of the singularity
in spherical dust collapse. We identify three different portions
of the singularity; ${\cal P}_0$, the end-point of the time-like
geodesic $T_0$ ($r=0, t<t_0$); ${\cal N}_0$ ($t=t_0,r=0$), the
visible portion of the singularity and ${\cal S}_0$
($t=t_c(r),r>0$), the censored portion of the singularity. $N_0$
is the unique ingoing radial null geodesic terminating at ${\cal
P}_0$.}
\end{figure}

\section{Structure of the singularity II: Some useful limits.}
In this section, we consider the limits of various quantities
along different causal geodesics terminating at and emanating from
the singularity.
\subsection{Causal geodesics running into the singularity}
We consider first the congruence of future-pointing time-like
geodesics $r=r_0\geq0$ with tangent ${\vec u}=
\frac{\partial}{\partial t}$. This is the congruence formed by the
world-lines of the dust particles and $t$ is proper time along
each member of the congruence. We study the approach to the
singularity from the point of view of the behaviour of the
characteristic lengths $l_\alpha$, $\alpha=1,2,3$ along the
eigen-directions of the expansion tensor
$\theta_{ab}=\nabla_au_b$, defined by \[ \frac{\dot
l_\alpha}{l_\alpha} = \theta_\alpha,\] where $\theta_\alpha$ is
the eigenvalue associated with the direction $\alpha$. The
geodesics in question have eigenvalues
\[\theta_1=\frac{R_{,rt}}{R_{,r}}=\frac13(t_c-t)^{-1}-(t_{sc}-t)^{-1}\]
and
\[\theta_2=\theta_3=\frac{R_{,t}}{R}=-\frac23(t_c-t)^{-1}.\]
Then the characteristic lengths $l_\alpha$ along $r=0$ satisfy (by
virtue of $t_c(0)=t_{sc}(0)=t_0$)
\[ l_1 = l_2 = l_3 = (t_0-t)^{2/3}.\]
Thus this singularity is a `point' in the Thorne classification
(see e.g. chapter one of \cite{dsincosmo}). On the other hand, the
characteristic lengths along $r=r_0>0$ satisfy
\begin{eqnarray*}l_1 &=&
(t_{sc}(r_0)-t)(t_c(r_0)-t)^{-1/3}\\&\sim&
(t_{sc}(r_0)-t_c(r_0))(t_c(r_0)-t)^{-1/3},\\
l_2&=&l_3 = (t_c(r_0)-t)^{2/3},\end{eqnarray*} so these
singularities (i.e.\ the {\em different} endpoints of these
geodesics) are `cigars' in this classification. Apart from the
issue of visibility, this is the first indication of a significant
distinction between the singularities ${\cal P}_0$ and ${\cal
S}_0$.

The shear and expansion scalars of this congruence are given by
\begin{eqnarray*}
\sigma^2 &=& \frac13((t_{sc}-t)^{-1}-(t_c-t)^{-1})^2,\\ \theta&=&
((t_{sc}-t)^{-1}+(t_c-t)^{-1}),
\end{eqnarray*}
while the density and Weyl invariant are
\begin{eqnarray*}
\rho&=&\frac43(t_c-t)^{-1}(t_{sc}-t)^{-1},\\
\Psi_2&=&\frac29(t_c-t)^{-2}(t_{sc}-t)^{-1}(t_c-t_{sc}).
\end{eqnarray*}

The limiting behaviour of ratios of these terms have significance
in the study of singularities, in particular in the context of the
Weyl curvature hypothesis and the study of isotropic singularities
which has grown up around this. See \cite{gw} for the ideas which
motivate this study, and \cite{se} for a more recent review of the
topic. The hypothesized scenario which these studies seek to
corroborate or deny is that the initial cosmological singularity has
low `gravitational entropy' \cite{penrose}
and consequently sub-dominant Weyl curvature, while
singularities arising from the end-state of collapse are highly
entropic and so are accompanied by dominant Weyl curvature.

We recall that isotropic singularities have been put forward as a model of the
initial cosmological singularity, and the sub-dominance of the Weyl curvature
is manifest in these models in that \be
W^2:=\frac{C_{abcd}C^{abcd}}{R_{ef}R^{ef}}\to 0\label{weylsd}\ee
in the limit as the singularity is approached \cite{gw}. Other
limits may also be derived for the case of isotropic singularities \cite{gw}, for example
\[
\lim\{\frac{A}{\theta^2},\frac{\sigma^2}{\theta^2},\theta\}=\{\frac13,0,+\infty\},\]
where $A=G_{ab}u^au^b$ and $u^a$ is tangent to a time-like
congruence. The limit is taken along this congruence, which is
assumed to have a certain degree of regularity.

Now, in the present case,
along the central world-line $T_0$, we find the following limits:
\[ \lim_{t\to
t_0}\{W^2,\frac{A}{\theta^2},\frac{\sigma^2}{\theta^2},\theta\}=\{0,\frac13,0,-\infty\},
\]
which reveals an interesting comparison with isotropic behaviour
summarised above.
In fact, the equality holds identically for the first three limits. The
sign difference in the last term arises from the fact that we are
studying collapse.

Along the world-lines $r=r_0>0$, the corresponding  limits are
\[ \lim_{t\to
t_c(r_0)}\{W^2,\frac{A}{\theta^2},\frac{\sigma^2}{\theta^2},\theta\}=\{+\infty,0,\frac13,-\infty\}.
\]
Note that the shear is
divergent in the approach to the singularity and that the first of these limits
indicates the dominance
of the Weyl curvature at ${\cal S}_0$. The latter
actually holds for any causal curve running into ${\cal S}_0$. To see
this we can write
\[ W = \frac16 \frac{t_{sc}-t_{c}}{t_c-t}.\]
Since $t_{sc}(r_0)>t_c(r_0)$ for $r_0>0$, this quantity diverges
to $+\infty$ in the approach to the singularity $t=t_c(r), r>0$
regardless of the direction of approach.

Again, we conclude that there is a significant
difference between the behaviour at ${\cal P}_0$ and at ${\cal
S}_0$.\\\\
As far as ingoing radial null geodesics are concerned it was
already demonstrated in section IV that there is a unique ingoing
radial null geodesic which terminates at ${\cal P}_0$. Repeating
the arguments of the next section (from (\ref{wlim}) onwards), we
can shown that the Weyl-to-Ricci ratio $W$ is finite in the limit
as the singularity is approached along this geodesic. This ratio
diverges in the approach to ${\cal S}_0$ along ingoing radial null
geodesics.
\subsection{Causal curves emerging from the singularity}
As we have shown above, the singularity is naked if and only if
there are time-like geodesics emerging the singularity.
These results only show existence of such geodesics and
give very little information about them, certainly not enough to
carry out the analysis done in the previous section. However, there
is one situation in which this analysis is possible, namely
self-similar collapse. This requires the use of a slightly
different gauge from the one used above. The line-element is
given by (\ref{eq1}), but the solution for $R$ is now \cite{dj92}
\[ R = r(1-\lambda x)^{2/3},\]
where $x=t/r$ and $\lambda$ is a constant parameter. Note that
$t=0$ is no longer a regular initial data surface, but any
$t=t_i<0$ is such. The shell-focussing singularity occurs at
$x=\lambda^{-1}$. The singularity $(r,t)=(0,0)$ is naked if and only if
$0<\lambda<\lambda_c\simeq 0.64$ \cite{dj92}. In the case where the
singularity is naked, the surfaces $x=$constant are space-like for
$x<x_0$ and for $x>x_1$, and time-like for $x_0<x<x_1$, where
$x_0,x_1$ are respectively the smallest and largest positive roots
of $x-R_{,r}(x)=0$. $x=x_0$ is the Cauchy horizon, i.e.\ is the
first outgoing radial null geodesic emerging from the singularity.
All outgoing radial null geodesics in the region $x_0\leq x \leq
x_1, r>0$ originate at the singularity $(0,0)$. Thus for
$x_c\in(x_0,x_1)$, the curves $t=x_cr$,$(\theta,\phi)=$constant
are time-like curves originating at the singularity. We analyse
this congruence under the same headings as the previous section.

Writing $R_{,r} = (1-\lambda x)^{-1/3}(1-\lambda x)=F(x)$, this
congruence has unit tangent vector field
\[ {\vec{u}} = \alpha(x) (x\frac{\partial}{\partial t}+\frac{\partial}{\partial
r}),\] where $\alpha(x)=(x^2-F^2)^{-1/2}$. The acceleration vector
field is
\[ {\vec{a}}=\frac{a(x)}{r}(-\frac{\partial}{\partial t}+x\frac{\partial}{\partial
r}),\] where \[ a(x) = -\frac29\lambda^2\alpha^2F(1-\lambda
x)^{-4/3}.\] The eigenvalues of the expansion tensor are found to
be all equal and are given by
\[ \theta_1=\frac{\alpha(x)}{r}.\]
Any given member of this congruence has, from the form of ${\vec
u}$ above,
\[ \frac{dr}{d\tau} = \alpha(x_c),\]
and so the proper time $\tau$ satisfies $\tau = r/\alpha(x_c)$. Thus
the characteristic lengths all satisfy
\[ \frac{{\dot{l}_\alpha}}{l_\alpha} =\frac{1}{\tau},\]
giving $l_\alpha = \tau$ for $\alpha=1,2,3$. Again we see that a
visible portion of the singularity consists of a singular region
which are `points' in Thorne's classification.

As the expansion tensor has three equal eigenvectors, the shear of
this congruence vanishes identically. The Weyl-to-Ricci ratio is
\[ |\frac{\Psi_2}{\rho}| = \frac13(1-\lambda x)^{-1},\]
which is non-zero and finite in the limit as the singularity is
approached along $x=x_c$. $\Psi_2$ and $\rho$ both individually
diverge at the singularity. The expansion $\theta = 3\alpha(x)/r$
diverges.

In the general (i.e.\ non-self similar case), we can analyze this
last limit for the time-like geodesics emerging from the
singularity whose existence has been proven in Proposition 3
above. The limit we wish to determine is
\begin{eqnarray*}W_0&=& \lim_{r\to 0}|\frac{\Psi_2}{\rho}|\\
&=&\lim_{r\to 0}\frac16 \frac{t_{sc}-t_c}{t_c-t},\end{eqnarray*}
where the limit is taken along the geodesic emerging from the
singularity. The value of this limit may be obtained
using l'Hopital's rule: \be W_0=
\lim_{r\to
0}\frac16\frac{t^\prime_{sc}-t^\prime_c}{t^\prime_c-t^\prime},\label{wlim}\ee
where $t^\prime$ is the slope of the geodesic in the $r-t$ plane.
Referring to the proof of Proposition 9 in \cite{mn1}, we recall that along such
geodesics, the slope $t^\prime(r)$ of these curves in the $r-t$
plane satisfies
\[ R_{,r}\leq t^\prime \leq (1+\delta)R_{,r},\]
for some $\delta>0$ and sufficiently small. The significance of
$R_{,r}$ is that it gives the slope of the outgoing radial null
geodesics and therefore the bounds above allow us to focus on the
radial null geodesics. We know that such geodesics emerge into the
region $t_0<t<t_H(r)$, where $t=t_H(r)=t_c-\frac23 m$ is the
apparent horizon. Using the latter bounds on $t$, we can obtain
bounds on $R_{,r}$ (see equation (\ref{rreq})), i.e.\ on
$t^\prime$. These in turn may be integrated to improve the
original bounds on $t$. Focussing on the case where
$m^\prime_1(0)\neq0$ and carrying out this process twice yields
bounds of the form \[ c_1r^{2/3}+u_1(r)\leq t^\prime\leq
c_2r^{2/3}+u_2(r),\] where $0<c_1<c_2$ and $u_i=o(r^{2/3})$ for
$i=1,2$. We can then take the limit in (\ref{wlim}) and find that
\[ W_0 =\lim_{r\to 0}
\frac16\frac{t^\prime_{sc}-t^\prime_c}{t_c^\prime-t^\prime} =\frac19.\]
In the case $m^\prime_1(0)=0,m^{\prime\prime}_1(0)\neq0$, the
limit gives $W_0=2/9$. These limits are the same for every causal
geodesic emerging from the singularity. Finally, for the case
$m^\prime_1(0)=m^{\prime\prime}_1(0)=0$,
$m^{\prime\prime\prime}_1(0)\neq 0$, the existence of the limit is
proved in the same way as above. In this case, the
result depends on the value of the
third derivative and on the geodesic in question.

Our principal conclusion is thus that the visible portion ${\cal
N}_0$ of the singularity does not exhibit the Weyl-dominance which
is thought to be characteristic of singularities arising in
collapse. Nor does it exhibit the Weyl-sub-dominance
(\ref{weylsd}) characteristic of isotropic singularities.   In the
case where the calculation may be carried out, the singularity is
point-like.

As an aside, we remark that we can find examples
where the quantity $W^2$ increases or decreases
with time depending on the geodesics under consideration. This is potentially of  interest
to debates on gravitational entropy (see e.g. \cite{penrose},
\cite{bonnor} and \cite{mt}) .
\subsection{Redshift}
In order to make contact with recent results given in \cite{desh} we
study briefly the redshift along outgoing geodesics.

Consider a photon which is emitted by a source with unit time-like
tangent $u^a_{(s)}$ and received by an observer with tangent
$u^a_{(o)}$. Let $k^a$ be the tangent to the photon's world-line. Then
the redshift $z$ relative to the emission and reception events $P_1$ and
$P_2$ is given by
\[ 1+z =\frac{[k_au^a_{(s)}]_{P_1}}{[k_au^a_{(o)}]_{P_2}}.\]
The redshift of photons emerging from a naked singularity has been
calculated in \cite{dwivedi98} and \cite{desh} for various
trajectories. The source and receiver are taken to be co-moving
with the fluid: take the source to be $r=r_0<<1$ and the receiver
to be $r=r_1>r_0$. Thus the source hugs the central world-line
$T_0$ and the null singularity ${\cal{N}}_0$ and terminates at
${\cal{S}}_0$ near to the junction of ${\cal{N}}_0$ and
${\cal{S}}_0$. Then $u^a_{(s)}=u^a_{(o)}=\delta^a_0$, and
$1+z={\dot{t}}_{P_1}/{\dot{t}}_{P_2}$. Taking $P_1$ to be the past
endpoint of a geodesic emerging into the future from the
singularity places the photon source on the singularity and can be
thought of as allowing $r_0\downarrow 0$. In \cite{dwivedi98}, it
was shown that the redshift thus measured is finite for
singularities of classes (i) and (ii) of section II, but infinite
for singularities of class (iii). In \cite{desh}, the redshift is
shown to be infinite for photons propagating along non-radial
geodesics emerging from the singularity. This latter result is
immediate as we see from (\ref{NRTLG3}): ${\dot{t}}$ diverges at
the singularity $R=0$ for any non-radial causal geodesic which is
not a radial null geodesic. This shows that the result - infinite
redshift -also applies to particle de Broglie wavelengths ($k^a$
time-like). This is a possible (classical) indication that
particles emerging from a naked singularity may not be associated
with the transfer of infinite amounts of energy. Furthermore, if
physical particles escaping from the singularity always travel
along geodesics with some angular momentum, then they will always
be infinitely redshifted to an outside observer. We emphasize that
these results are purely classical and it would be interesting to
study them using e.g. a semi-classical approximation.
\section{Structure of the singularity III: Topology}
In the previous section, we classified the different portions of
the singularity using Thorne's point-cigar-barrel-pancake
classification. This describes the behaviour of a co-moving fluid
element carried by a time-like congruence meeting the singularity.
However, this does not address another question which relates to
the topology of the singularity and which we shall now describe.
In the
conformal diagram of Figure 1, each point in the two-dimensional
representation of the non-singular part of the space-time manifold
represents a 2-sphere: the point $(r,t)$ of this diagram
represents the 2-sphere with radius $R=R(r,t)$ at time $t$. The
question we address here is the following: does each point on the singular
boundary ${\cal P}_0\cup{\cal N}_0\cup{\cal S}_0$ represent topologically a
2-sphere or a point? Christodoulou has shown that the future
outgoing null (and hence censored) singular boundary which can
arise in the collapse of a self-similar scalar field is foliated by
points, while the space-like singular boundary that arises in
another sector of the space of solutions is foliated by 2-spheres
\cite{christo-scalar}.

Christodoulou's argument runs as follows.
Any radial null geodesic of the space-time actually represents a
2-sphere's worth of null geodesics. Consider a radial null
geodesic $\gamma$ running into a future outgoing null singular
boundary ${\cal N}_1$. This actually represents a family
$\gamma^\xi, \xi\in\bf{S}^2$ of radial null geodesics running
into the singularity. From this family, choose a fixed member
$\gamma^N$ and the antipodal member $\gamma^S$ ($N,S$ for north, south).
Now let $\{p^{N,S}_n\}$ be causally increasing sequences of points
along $\gamma^{N,S}$ respectively, chosen so that the
singularity is reached in the limit $n\to\infty$ and for
each $n$, the points $p^{N,S}_{n}$ are antipodal on the same 2-sphere
$\bf{S}^2_n$ . Christodoulou shows that  the causal pasts
$J^-(p^{N,S}_n)$ of $p^{N,S}_n$ satisfy \[\lim_{n\to\infty}
J^-(p^N_n)=\lim_{n\to\infty}J^-(p^S_n),\] and so the ideal
boundary points $\lim_{n\to\infty} p^{N,S}_n$, having the same
causal past, are, by definition, identical. See Figure 2.

\begin{figure}[!htb]
\centerline{\def\epsfsize#1#2{1 #1}\epsffile{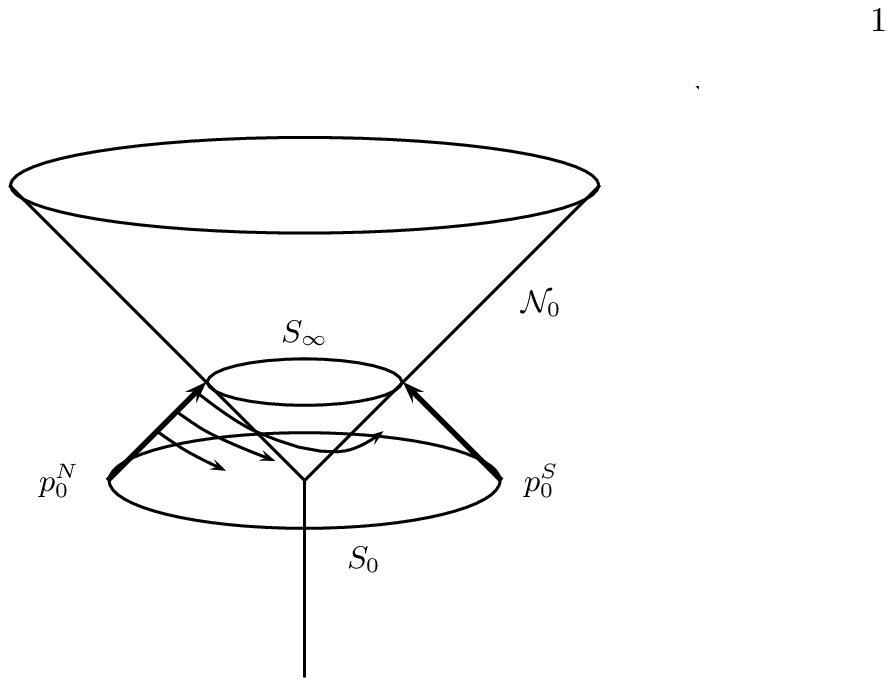}}
\caption{\label{figure2} Behaviour of non-radial geodesics near
the singularity. For convenience, we consider a future outgoing
null singular boundary ${\cal N}_0$. $R=0$ along ${\cal N}_0$. The
straight arrows represent ingoing radial null geodesics
$\gamma^{N,S}$ with initial points $p^{N,S}_0$ which are antipodal
points on an initial 2-sphere $S_0$. These geodesics project to
the same radial null geodesic in the Lorentzian 2-space
$d\Omega=0$.  The curved arrows are past-directed non-radial null
geodesics emerging from different points  $p^N_n\in\gamma^N$, the
different lengths of these curves representing the fact that
angular velocity ${\dot\phi}$ of these geodesics increases as
their initial point $p^N_n$ approaches the singularity $R=0$. In
the case where $\phi\to-\infty$ as the singularity is approached,
the geodesics reach all the way around the regular center $R=0$ to
the opposite side, and in the limit, we may consider that the
non-radial geodesic emitted from $p^N_\infty$ reaches
$p^S_\infty$. Hence as points of the causal boundary, $p^N_\infty$
and $p^S_\infty$ are identified and $S_\infty$ is a single point.
In the case where $\phi$ has a finite limit as the singularity is
approached, $S_\infty$ is a 2-sphere.}
\end{figure}

The key ingredient required for this argument is to show
that past-directed non-radial null geodesics through $p^N_n$ orbit
the axis $R=0$ sufficiently quickly (in the limit as $n\to\infty$)
so that the causal past of $p^S_n$ is contained in the causal past of
$p^N_n$. That is, the rapidly orbiting non-radial null geodesics
have enough `time' to get around from the north to the south pole
of $\bf{S}^2_n$ and so have a look at what is to be seen from
there. See Figure 2.

We argue here that, at least in the self-similar case, the
topology of the naked portion of the singularity is not uniform,
being point-like on one region and spherical on another. This
requires a fairly thorough analysis of the behaviour of null
geodesics emerging from the singularity.\\\\
In the self-similar case,
the line element may be written (see e.g. \cite{dj92}) \be ds^2 = -dt^2
+F^2(x)dr^2+r^2U^2(x)d\Omega^2,\label{lelss}\ee where $x=t/r$ and
\[ U(x)  = (1-\lambda x)^{2/3},\qquad F(x)=(1-\lambda
x)^{-1/3}(1-\frac{\lambda}{3}x).\] The shell-focussing singularity
occurs at $x=\lambda^{-1}$.   The equations governing outgoing
radial null geodesics (ORNG's) may be written as
\[ \frac{dx}{dr} =\frac{F-x}{r}.\]
Then it can be shown that the singularity is naked for $\lambda$
in the range $(0,\lambda_c)$, in which case $F-x$ has two positive
roots $x_0<x_1<\lambda^{-1}$. The curves $x=x_0$ and $x=x_1$ are outgoing
radial null geodesics emerging from the singularity $(t,r)=(0,0)$.
We recall that $x_0$ corresponds to the Cauchy horizon, i.e.\ the boundary of the future of a
regular initial data surface $t=t_i<0$.

ORNG's initially in the region $x\in(x_0,x_1)$ remain in this
region and satisfy $x^\prime(r)=(F-x)/r<0$. Thus along such a
geodesic, $x$ increases as $r\downarrow0$. But $x$ is bounded
above by $x_1$, and so must have a finite limit as $r\downarrow0$.
This limit $(\lim\frac{t}{r}=\lim\frac{dt}{dr})$ must solve $x=F$
and so must equal $x_1$. A similar argument applies for ORNG's in
$x\in(x_1,\lambda^{-1})$. Hence all ORNG's  emerging from the
singularity, except for the single ORNG $x\equiv x_0$, originate
at $(0,0)$ with $x=x_1$. However as we will see, non-radial null
geodesics may emerge with different values of $x$ in this limit.
We note that this indicates that considering only radial null
geodesics does not give a complete picture of the singular
boundary.\\\\
We now consider the question raised above. Each of the radial null
geodesics emerging from $(0,0)$ represents a 2-sphere's worth of
ORNG's emerging from the singularity. To address the issue of the
topology of the singularity, we consider non-radial null geodesics
(NRNG's) emerging from the singularity.

The homothetic Killing vector field which generates the
self-similarity of (\ref{lelss}) is
\[ \vec{\xi}=t\frac{\partial}{\partial t}+r\frac{\partial}{\partial
r}.\] If $k^a$ is tangent to a null geodesic, then
$g_{ab}\xi^ak^b$ is constant along the geodesic. This is a first
integral of the null geodesic equations, which we may therefore
write as
\begin{eqnarray}
-{\dot t}^2+F^2{\dot r}^2+\frac{L^2}{r^2U^2}&=&0,\label{nrng1}\\
-{\dot t}+\frac{F^2}{x}{\dot r}+\frac{k}{rx}&=&0,\label{nrng2}
\end{eqnarray}
where $L$ is the conserved angular momentum and $k$ is the
constant generated by $\vec{\xi}$. We can assume that $k\geq 0$,
as a change of sign in $k$ corresponds to a change in sign of the
affine parameter. Our principal result which is of significance
for the topology of the singularity is the following:

\begin{prop}
(i) Through every point of $\{(t,r):x_0<x<x_1\}$, there exists a
non-radial null geodesic which emerges from the singularity
$(0,0)$. Along this geodesic, ${\dot\phi}$ is a non-integrable
function of the affine parameter $\tau$, and $\phi$ diverges to
$-\infty$ in the limit as the singularity is approached.
\newline\noindent(ii) Through every point of $\{(t,r):x_1<x<\lambda^{-1}\}$,
there exist non-radial null geodesics which emerge from the
singularity $(0,0)$. Along every such geodesic, $\phi$ has a
finite limit as the singularity is approached.
\end{prop}

The proof is given in the appendix. Proposition 5 allows us to
deduce the following results regarding the topological
nature of the
singularity ${\cal{N}}_0$, from the causal boundary perspective.

We start by recalling that radial null geodesics emerge into the
region $x_0<x<x_1$. Now, on any
such geodesics, we may choose a point $p^N$ sufficiently close to
the singularity such that there exists a non-radial null geodesic
through $p^N$ with the property than this NRNG reaches the antipodal
point of the 2-sphere (on which $p^N$ lies) before the radial geodesic
reaches the singularity. The significance of this is that in the
limit as the point $p^N$ approaches the singularity, this point
and its antipodal partner $p^S$ must be considered to have the
same causal future, and so are identified as points of the causal
boundary. Therefore,
the corresponding section of ${\cal{N}}_0$ is foliated by points.

On the other hand, the results obtained for NRNG's in the region
$x_1<x<\lambda^{-1}$ indicate that the NRNG's through a point
$p^N$ {\em do not} have enough time to reach the antipodal point
$p^S$ before the singularity is reached. Hence in the limit as the
singularity is approached, the causal futures of $p^N$ and $p^S$
{\em will not} match up with one another, and hence these points
cannot be identified as points of the causal boundary. In fact
$p^N$ will not have the same limiting causal future as {\em any}
other point on the same 2-sphere, and so the corresponding portion of ${\cal
N}_0$ is foliated by 2-spheres.

Finally, one further point of relevance may be deduced from the proof of
Proposition 5: In every case considered, the future evolution of
the NRNG's terminates at $x=\lambda^{-1}$ at some $r>0$. The
limiting behaviour of ${\dot r}$ and ${\dot x}$ in terms of $x$
can be read off from the governing equations, and can be used to
obtain the limiting behaviour of ${\dot R}={\dot r}U+rU^\prime
{\dot x}$. This can be integrated, and generically yields a result
indicating integrability of $1/R^2$. Hence $\phi$ has a
finite limit along any NRNG approaching the singularity
$x=\lambda^{-1}$ at some $r>0$. Thus the singularity ${\cal S}_0$
is foliated by 2-spheres.
\section{Conclusions and comments}
We have  extended the result of \cite{mn1} to all f.p.\ causal
geodesics by proving that for the marginally bound spherical dust
collapse there are radial and non-radial timelike geodesics which
emanate from the ensuing singularity if and only if there are
radial null geodesics which emanate from the singularity. It
remains to be seen whether this result is true in general in
spherical symmetry, at least for singularities at zero radius. The
fact that the naked singularity always has time-like geodesics
emerging therefrom leads to interesting possibilities in terms of
the spectrum of particles that may be created by such a
singularity: such particles may be massive as well as massless.
This would seem to be vital if the information lost in the process
of collapse is to re-emerge at a later stage in the form of a
black hole or naked singularity explosion \cite{Harada-etal00}.

We have studied the structure of the singularity from several
points of view. We started by studying the behaviour of limits of
various quantities along causal geodesics terminating at the
singularity and emanating from the singularity. In this way we
have classified different sectors of the singularity as points and
cigars. We have then considered the topological properties of the
singularity. These studies were motivated by the fact that the
existence of non-radial null geodesics emerging from the
singularity suggests that it must have some non-pointlike
structure. Naively, one might expect that the singularity ${\cal
N}_0$ of Figure 1 must by cylindrical, i.e.\ foliated by
2-spheres, since points thereon emit a full 2-spheres worth of
non-radial null geodesics. This is in contrast to a regular center
of a spherically symmetric space-time, through which non-radial
null geodesics cannot pass. However detailed calculations indicate
that while the topology of ${\cal S}_0$ is that of a 2-sphere, the
topology of ${\cal N}_0$ (in the sense of ideal boundary points)
may be either that of a 2-sphere or of a point.

The calculations in section V of various limits at the singularity
seem to us to be of interest beyond merely classifying the
different portions of the singularity. The question of what
constitutes a genuine space-time singularity is not yet settled.
This is of particular relevance for cosmic censorship hypothesis:
a visible singularity which is mild in some suitably (and
rigorously) defined way should not be considered a counter-example
to the hypothesis. The following two definitions have proved to be
useful. Both involve the question of predictability to the future
of a singularity.

We mention first that of Clarke \cite{clarke}. Put simply, a
singularity is considered {\em inessential} if it does not present
any obstruction to the evolution of test fields in space-time. The
question of global hyperbolicity of the space-time (cosmic
censorship) is redirected to the question of generalized
hyperbolicity of fields on the space-time. Examples of space-times
admitting singularities in the classical sense of $C^{2-}$
geodesic incompleteness but satisfying a generalized hyperbolicity
condition have been given in \cite{clarke}, \cite{vic-wil1} and
\cite{vic-wil2}.

The second definition is that of an isotropic singularity
\cite{gw}. As mentioned above, this was introduced in the study of
the initial cosmological singularity and hypotheses related to
this \cite{penrose}. The connection with cosmic censorship is
perhaps not immediately obvious, but becomes so when one
acknowledges that the initial cosmological singularity is naked:
every past-directed causal geodesic meets the singularity. The
isotropic singularity program has produced the following profound
result: the big bang can be considered a regular initial data
surface for the Einstein field equations. (Different matter models
have been studied; polytropic perfect fluids in \cite{ang-tod1},
the Einstein-Vlasov model in \cite{ang-tod2}. These rely on
results for singular differential operators due to Newman-Claudel
\cite{new-claud}.) Thus with isotropic singularities, we see that
it is possible to have global hyperbolicity {\em of the
space-time} to the future of a singularity.

So with a view to addressing the question of whether or not the
naked singularities studied here should be considered genuine
counter-examples to cosmic censorship we could ask: do we have any
evidence that the singularity is either (i) inessential in
Clarke's sense or (ii) isotropic? We will not address the first of
these here. Regarding the second, we contend that there is some
evidence that, at least in the self-similar case, the singularity
${\cal{N}}_0\cup{\cal{P}}_0$ has characteristics in common with
isotropic singularities.

As we have seen, the Weyl-sub-dominance property (\ref{weylsd}) -
a characteristic of isotropic singularities -  is {\em not}
satisfied for the congruence of time-like curves emerging from the
singularity which we studied in Section 5.2. Nevertheless, we
speculate that the fact that the ratio $W^2$ is finite (but not
zero) may turn out to be an indication of good behaviour of ${\cal
N}_0$ regarded as an initial data surface. As evidence in favour
of this, consider the following details which relate again to the
self-similar case. From the calculations in the appendix, we know
that outgoing radial null geodesics emerging from the singularity
fall into two classes. The first class contains a single member,
the ORNG $x=x_0$, which emerges from ${\cal P}_0$ in Figure 1. The
second class consists of all other ORNG's emerging from the
singularity: these emerge from ${\cal N}_0$. Along these
geodesics, the limit $x\to x_1$ is satisfied as the singularity is
approached. Hence the congruence of time-like curves
$x=x_c\in(x_0,x_1)$ - call it the similarity congruence -
considered in \S V.B must be considered to emerge from ${\cal
P}_0$. This resembles the behaviour of the initial singularity in
certain FLRW universes, namely those filled with a perfect fluid
satisfying a equation of state $p=\alpha\rho$, with
$\alpha\in(-1,-1/3]$. In these space-times, the entire fluid
congruence emerges from a single point of the singularity and a
null portion of the singularity forms a portion of the causal
future of this point. See \cite{seno} for details. This
resemblance is borne out by writing the line-element in
coordinates adapted to the similarity congruence. Introduce a time
function $T=rH(x)$ chosen so that the similarity congruence is
orthogonal to $T=$constant. Then the line-element becomes
\[ ds^2 =
\frac{r^2}{(x^2-F^2)}\left\{-(x^2-F^2)^2\frac{dT^2}{T^2}+F^2dx^2+U^2
d\Omega^2\right\}.\] This applies for $r>0$ and $x\in(x_0,x_1)$,
wherein $x^2-F^2>0$. The function $H$ satisfies
\[ \frac{H^\prime}{H}=\frac{x}{x^2-F^2},\]
and therefore is finite on $(x_0,x_1)$. The line-element may be
written as $ds^2=T^2 d\bar s^2$, where
\[ d\bar s^2 = -a^2(x)\frac{dT^2}{T^2}+b^2(x)dx^2+c^2(x)d\Omega^2,\]
with $a,b,c$ finite and non-zero on $(x_0,x_1)$. $T$ is the
exponential of proper time along the similarity congruence. This
decomposition of the metric into an overall scaling factor (which
vanishes at the singularity) times a non-singular metric (when
written in terms of $\tau=\ln T$) is exactly the same
decomposition which exists for the FLRW metrics mentioned above,
and is similar to the decomposition whose existence defines
isotropic singularities (the difference is that $\tau\to-\infty$
at the singularity currently under discussion).

So while there is still no direct evidence that this naked
singularity does not break the future predictability of the
space-time, the question appears to be worthy of future
investigation.

\section*{Acknowledgements}
We thank T. Harada, M. MacCallum, J. Senovilla and R. Tavakol for
interesting discussions. BCN acknowledges support from DCU under
the Albert College Fellowship Scheme 2001, and thanks the
Relativity Group at Queen Mary, University of London, the
Department of Theoretical Physics, University of the Basque
Country and the Department of Mathematics, University of Minho for
hospitality. FCM thanks CMAT, Univ. Minho, for support and FCT
(Portugal) for grant PRAXIS XXI BD/16012/98.

\appendix
\section{Proof of proposition 5}

(i) Given $x_c\in(x_0,x_1)$, it is straightforward to show that
there are NRNG's confined to this time-like hypersurface which
have the desired properties. Substituting $x=$constant into
(\ref{nrng1}) and (\ref{nrng2}) yields
\[ r{\dot r} =\pm\frac{L}{U}(x^2-F^2)^{-1/2}=k(x^2-F^2)^{-1},\]
which is valid since $x^2>F^2$ in $(x_0,x_1)$. Choosing $k,L$ so
that these two match, which gives a 1-parameter solution, gives
the geodesic in question. A sign choice is required in order that
the geodesic is future pointing ${\dot t}>0$. Note that $r{\dot
r}=c$, which is positive for the future-pointing choice and so
$r^2\sim 2c\tau$ as $\tau\downarrow0$, choosing the origin of the
affine parameter $\tau$ to coincide with the singularity. Then
\[{\dot \phi}\sim \frac{L}{2c}\tau^{-1},\]
and so $\phi$ diverges to $-\infty$ as the singularity is
approached.

\noindent(ii) To prove this part of the proposition requires the
study of {\em all} non-radial null geodesics in the region
$(x_1,\lambda^{-1})$. The equations (\ref{nrng1}) and
(\ref{nrng2}) yield a quadratic equation for ${\dot r}$, which has
solutions given by
\be r{\dot
r}=\frac{1}{F^2-x^2}[-k\pm\frac{x}{F}(k^2+\frac{L^2}{U^2}(F^2-x^2))^{1/2}].\label{rdeq}\ee
These give \be {\dot x}
=\pm\frac{1}{r^2F}(k^2+\frac{L^2}{U^2}(F^2-x^2))^{1/2}.\label{xdeq}\ee

The solutions with the lower sign require the change
$\tau\to-\tau$ in order to be explicitly future-pointing.
Implementing this, the equations are (any two of)
\begin{eqnarray}
r{\dot
r}&=&\frac{1}{F^2-x^2}[k+\frac{x}{F}(k^2+\frac{L^2}{U^2}(F^2-x^2))^{\frac{1}{2}}]>0,
\label{lrreq}\\
{\dot x}
&=&\frac{1}{r^2F}(k^2+\frac{L^2}{U^2}(F^2-x^2))^{1/2}>0,\label{lrxeq}\\
{\dot t}&=&\frac{F^2}{x}{\dot r}-\frac{k}{rx}>0.\label{lrteq}
\end{eqnarray}
Recall that $k\geq 0$. Positivity of ${\dot x}$ immediately gives
\be t<x_i r\label{bdd1}\ee for any point $(t,r)$ preceding a
chosen initial point $(t_i,r_i)$. Positivity of ${\dot r}$ and $k$
yields
\begin{eqnarray*} r\frac{dx}{dr}&<&\frac{F^2-x^2}{x}\\
&<& 2(F^\prime_1-1)(x-x_1)+M(x-x_1)^2.
\end{eqnarray*}
There latter inequality arises from using Taylor's theorem on $F$,
which is analytic at $x_1$. $2M>0$ is a bound on the closed
interval $[x_1,x_i]$ of the (continuous) second derivative of
$(F^2-x^2)/x$. $F^\prime_1=F^\prime(x_1)$, and this number is
always bigger than 1. Integrating over $[r,r_i]$, where $0<r<r_i$
yields \be t>x_1r+(\frac{c_1}{1-c_2r^A})r^{A+1},\label{bdd2}\ee
where $c_{1,2}$ are positive constants and $A=2(F^\prime_1-1)>0$.
Combining (\ref{bdd1}) and (\ref{bdd2}), we see that these
geodesics all originate at $(t=0,r=0)$.

The rate at which $r$ shrinks to zero can be obtained as follows.
We have
\[ r{\dot r}\sim \frac{2k}{F^2-x^2}\sim
\frac{k}{x_1(F^\prime_1-1)}(x-x_1)^{-1},\] while
\[ {\dot x}\sim\frac{k}{x_1}r^{-2}.\]
These may be integrated to give
\[ r\sim k\tau^{1/(F^\prime_1+1)}=k\tau^\alpha\]
where $\alpha<1/2$ and so
\[ \phi(\tau) = \int_0^\tau \frac{L}{R^2(\nu)}\,d\nu\]
converges to a finite value in the limit as $\tau\downarrow0$.

Next we consider the solutions corresponding to the choice of
upper sign in (\ref{rdeq}) and (\ref{xdeq}). We note that ${\dot
t}>0$ and ${\dot x}>0$ are automatic consequences of this choice.
On the other hand, ${\dot r}>0$ if and only if $H(x)>0$, where
\[ H=\frac{L^2}{U^2}-\frac{k^2}{x^2}.\] We note that
$H^\prime(x)>0$. Then for any initial value $H(\tau_0)$ of $H$,
the increase of $x$ brings about the increase of $H$ to positive
values. Then ${\dot r}$ becomes and remains positive, and so the
geodesic must run into the singularity $x=\lambda^{-1}$. The past
evolution requires more careful attention. We deal with the case
$k>0$ first.

We have $r{\dot r}=r{\dot r}(x)$, and by using a Taylor expansion
centered on $x_1$, we find
\[ r{\dot r}(x) =\frac{H(x_1)}{2k}+O(x-x_1).\]
If in the past evolution we reach $\tau_1<\tau_0$ whereat
$H(\tau_1)<0$ and $x(\tau_1)>x_1$, then $H<0$ for all
$\tau<\tau_1$. Hence $r$ increases into the past, and so the
geodesic inevitably extends back to $x=x_1$ at some $r>0$, thus
avoiding the singularity. This situation is obtained
if and only if
$H(x_1)<0$.

So it remains to consider the cases $H(x_1)\geq 0$. If
$H(x_1)>0$, then the argument above shows that $r{\dot r}>0$ for
all $x>x_1$.

We can obtain the bound $rx^\prime(r)>(F^2-x^2)/x$. Integrating
over $[r,r_i]$ ($0<r<r_i$) yields
\[ \int_r^{r_i} \frac{x}{F^2-x^2}\,dx>\ln|\frac{r_i}{r}|.\]
Thus if the geodesic approaches $r=0$ in the past, then the
integral on the left hand side here must diverge. This can only
occur if we also have $x\downarrow x_1$ as the singularity is
approached. Let this occur at $\tau=\tau_s$. But then
\begin{eqnarray*}
x_1&=&\lim_{\tau\to\tau_s}\frac{t}{r}\\
&=&\lim_{r\to 0}\frac{dt}{dr}\\
&=&\lim_{x\to x_1}(\frac{F^2}{x}+\frac{k}{x r{\dot r}})\\
&=& x_1 +
2\frac{k^2}{x_1}(\frac{L^2}{U_1^2}-\frac{k^2}{x_1^2})^{-1},
\end{eqnarray*}
which yields a contradiction. Therefore these geodesics must
extend back to $x=x_1$ before reaching $r=0$. This argument
applies when $H(x_1)>0$ and requires only slight modification to
be applied to the case $H(x_1)=0$. Note that there is a finite
(rather than infinitesmal) interval of affine parameter between an
initial point on the geodesic and the point corresponding to entry
into the region $x\leq x_1$.

Finally, we need to consider the NRNG's with $k=0$. Note that in
this case, the choice of sign in (\ref{rdeq}) and (\ref{xdeq}) is
only of a time-orientation and so without loss of generality
we only treat
the choice of upper sign.  The argument leading up to (\ref{bdd2})
may be repeated to show that in this case, there exist geodesics
emerging from $(0,0)$, and that $x\to x_1$ along these geodesics
as the singularity is approached. It remains to determine the rate
at which $r$ approaches zero for these geodesics.

From the governing equations, we have ${\dot x}=G(x)r^{-2}$, where
\[ G(x) =\frac{L}{FU}(F^2-x^2)^{1/2}.\]
We can also write
\[ (r^2)^\cdot=2r^2x(F^2-x^2)^{-1}{\dot x}.\]
These yield the second order equation
\[{\ddot x}=J(x){\dot x}^2,\]
where
\begin{eqnarray*}J(x)&=&(\frac{G^\prime}{G}-\frac{2x}{F^2-x^2})\\
&=&\frac12(\frac{F^\prime_1-3}{F^\prime_1-1})(x-x_1)^{-1}+O(1)\quad
\hbox{{\rm as}}\quad x\to x_1. \end{eqnarray*}  This equation may
be integrated to obtain the $\tau-$dependence of $x$ as
$x\downarrow x_1$. Then using $r^2 = H/{\dot x}$, we obtain
\[ r^2\sim c^2 \tau^{2/(F^\prime_1+1)}\quad \hbox{{\rm as}}\quad r\to
0.\] Recalling that $F^\prime_1>1$, we see that the power here is
less than one. Hence \[ {\dot
\phi}=\frac{L}{R^2}\sim\frac{L}{U^2(x_1)r^2}\] is integrable in
the limit as the singularity is approached, giving a finite limit
for the value of $\phi$.

To summarise: NRNG's in $x>x_1$ either exit this region at some
finite radius, taking a finite amount of time to do so, or emerge
from the naked singularity $(0,0)$. In the latter case, $\phi$ has
a finite limit in the approach to the singularity. This concludes
the proof of the proposition.



\begin{thebibliography}{12}
\bibitem[1]{sj}  Singh T P \& Joshi P S 1996 {\em
Class. Quantum Grav.} {\bf 13} 559
\bibitem[2]{mn1} Mena F C \& Nolan B C 2001 {\em Class. Quantum Grav.} {\bf 18}
4531.
\bibitem[3]{desh} Deshingkar S S, Joshi P S \& Dwivedi I W, 2001. {\em Appearance of the central singularity in spherical collapse}.
To appear in {\em Phys. Rev. D} Published electronically as
gr-qc/0111053.
\bibitem[4]{seno} Senovilla J M M, 1998 {\em Gen. Rel. Grav.} {\bf 30}  701
\bibitem[5]{dsincosmo} Wainwright J \& Ellis G F R (eds.) 1997 {\em
Dynamical Systems in Cosmology} (Cambridge: Cambridge university
Press)
\bibitem[6]{christo-scalar}Christodoulou D, 1994 {\em Ann. Math.} {\bf 140} 607.
\bibitem[7]{scott} Scott S \& Szekeres P, 1986 {Gen. Rel. Grav.}, {\bf 18} 557
\bibitem[8]{penrose} Penrose R, in {\em General Relativity: An
Einstein Centenary Survey}, eds. Hawking S W \& Israel W, (Cambridge:
Cambridge University Press, 1979)
\bibitem[9]{gw} Goode S W \& Wainwright J 1985 {\em Class.
Quantum Grav.}, {\bf 2} 99.
\bibitem[10]{Lemaitre} Lemaitre G 1933 {\em Ann. Soc. Sci. Bruxelles}, {\bf
A53} 51
\bibitem[11]{Tolman} Tolman R C 1934 {\em Proc. Nat. Acad. Sci.}, {\bf 20}
\bibitem[12]{Bondi} Bondi H 1947 {\em Mon. Not. R. Astron. Soc.}, {\bf 107} 410
\bibitem[13]{eardley-smarr}
Eardley D M \& Smarr L 1979 {\em Phys. Rev.} D{\bf 19} 2239
\bibitem[14]{christo-dust}
Christodoulou D 1984 {\em Commun. Math. Phys.} {\bf 93}
171\bibitem[15]{newman} Newman R P A C 1986 {\em Class. Quantum
Grav.} {\bf 3} 527
\bibitem[16]{nolan99}
Nolan B C 1999 {\em Phys. Rev.} D{\bf 60} 024014
\bibitem[17]{nmg} Nolan B C, Mena F C \& Gon\c{c}alves S M C V
2002 {\em Phys. Lett. A}  {\bf 294}, 122.
\bibitem[18]{se} Scott S M \& Ericksson G 1998 in {\em Current
Topics in Mathematical Cosmology} ed. M Rainer and H-J Schmidt
(Singapore: World Scientific)
\bibitem[19]{dj92} Dwidevi I H \& Joshi P S 1992 {\em Commun. Math. Phys.} {\bf 146} 333
\bibitem[20]{bonnor} Bonnor W B 1985 {\em Phys. Lett. A} {\bf 112} 26
\bibitem[21]{mt} Mena F C \& Tavakol R 1999 {\em Class. Quant. Grav.}
{\bf 16} 435
\bibitem[22]{dwivedi98} Dwivedi I H 1998 {\em Phys. Rev. D} {\bf 58} 064004
\bibitem[23]{Harada-etal00} Harada T, Iguchi H \& Nakao K, 2000 {\em Phys.Rev. D} {\bf 61} 101502; {\em ibid.}
{\bf 62} 084037.
\bibitem[24]{clarke} Clarke C J S 1998 {\em Class. Quantum Grav.}
{\bf 15} 975.
\bibitem[25]{vic-wil1} Vickers J A  \& Wilson J P 2000 {\em Class. Quantum Grav.} {\bf 17}
1333.
\bibitem[26]{vic-wil2} Vickers J A \& Wilson J P 2001 {\em Generalised hyperbolicity: hypersurface
singularities} University of Southampton preprint. Published
electronically as gr-qc/0101018.
\bibitem[27]{ang-tod1}
Anguige K \& Tod K P  1999 {\em Ann Phys. (NY)} {\bf 276}, 257.
\bibitem[28]{ang-tod2}
Anguige K \& Tod K P  1999 {\em Ann Phys. (NY)} {\bf 276}, 294.
\bibitem[29]{new-claud}
Claudel C M \& Newman K P 1998 {\em Proc. Roy. Soc. Lond.} {\bf
454}, 1073.

%
%


%

\end{thebibliography}
\end{document}